# Title: Synchronous X-ray and Radio Mode Switches: a Rapid Global Transformation of the Pulsar Magnetosphere

**Authors:** W. Hermsen[1,2,*], J.W.T. Hessels[3,2], L. Kuiper[1], J. van Leeuwen[3,2], D. Mitra[4], J. de Plaa[1], J. M. Rankin[5,2], B. W. Stappers[6], G.A.E. Wright[7], R. Basu[4], A. Alexov[8], T. Coenen[2], J.-M. Grießmeier[9,14], T.E. Hassall[10,6], A. Karastergiou[11], E. Keane[12], V.I. Kondratiev[3,13], M. Kramer[12,6], M. Kuniyoshi[12], A. Noutsos[12], M. Serylak[14,9], M. Pilia[3], C. Sobey[12], P. Weltevrede[6], K. Zagkouris[11], A. Asgekar[3], I.M. Avruch[1,15,3], F. Batejat[16], M.E. Bell[17,10], M.R. Bell[18], M.J. Bentum[3,19], G. Bernardi[20], P. Best[21], L. Bîrzan[22], A. Bonafede[23], F. Breitling[24], J. Broderick[10], M. Brüggen[23], H.R. Butcher[3,25], B. Ciardi[18], S. Duscha[3], J. Eislöffel[26], H. Falcke[27,3,12], R. Fender[10], C. Ferrari[28], W. Frieswijk[3], M.A. Garrett[3,22], F. de Gasperin[23], E. de Geus[3], A.W. Gunst[3], G. Heald[3], M. Hoeft[26], A. Horneffer[12], M. Iacobelli[22], G. Kuper[3], P. Maat[3], G. Macario[28], S. Markoff[2], J.P. McKean[3], M. Mevius[3,15], J.C.A. Miller-Jones[29,2], R. Morganti[3,15], H. Munk[3], E. Orrú[3,27], H. Paas[30], M. Pandey-Pommier[22,31], V.N. Pandey[32], R. Pizzo[3], A.G. Polatidis[3], S. Rawlings[11], W. Reich[12], H. Röttgering[22], A.M.M. Scaife[10], A. Schoenmakers[3], A. Shulevski[15], J. Sluman[3], M. Steinmetz[24], M. Tagger[9], Y. Tang[3], C. Tasse[33,34,35], S. ter Veen[27], R. Vermeulen[3], R.H. van de Brink[3], R.J. van Weeren[20,22,3], R.A.M.J. Wijers[2], M.W. Wise[3,2], O. Wucknitz[36,12], S. Yatawatta[3], P. Zarka[33]

Affiliations:

[1]SRON, Netherlands Institute for Space Research, Sorbonnelaan 2, 3584 CA, Utrecht, The Netherlands.

[2]Astronomical Institute "Anton Pannekoek", University of Amsterdam, Postbus 94249, 1090 GE, Amsterdam, The Netherlands.

[3]ASTRON, the Netherlands Institute for Radio Astronomy, Postbus 2, 7990 AA Dwingeloo, The Netherlands.

[4]National Centre for Radio Astrophysics (NCFRA-TIFR), Post Bag 3, Ganeshkhind, Pune University Campus, Pune 411007, India.

[5]Physics Department, University of Vermont, Burlington, VT 05405, USA.

[6]Jodrell Bank Center for Astrophysics, School of Physics and Astronomy, The University of Manchester, Manchester M13 9PL, UK.

[7]Astronomy Centre, University of Sussex, Falmer, Brighton BN1 9QJ, UK.

[8] Space Telescope Science Institute (STScI), 3700 San Martin Drive, Baltimore, MD 21218, USA

[9] Laboratoire de Physique et Chimie de l' Environnement et de l' Espace, LPC2E CNRS/Université d'Orléans, 45071 Orléans Cedex 02, France.

[10] School of Physics and Astronomy, University of Southampton, Southampton, SO17 1BJ, UK.

[11]Astrophysics, University of Oxford, Denys Wilkinson Building, Keble Road, Oxford OX1 3RH , UK.

[12]Max-Planck-Institut für Radioastronomie, Auf dem Hügel 69, 53121 Bonn, Germany.

[13] Astro Space Center of the Lebedev Physical Institute, Profsoyuznaya str. 84/32, Moscow 117997, Russia2

[14] Station de Radioastronomie de Nançay, Observatoire de Paris, CNRS/INSU, 18330 Nançay, France.

[15] Kapteyn Astronomical Institute, PO Box 800, 9700 AV Groningen, The Netherlands.

[16] Onsala Space Observatory, Dept. of Earth and Space Sciences, Chalmers University of Technology, SE-43992 Onsala, Sweden.

[17] ARC Centre of Excellence for All-sky astrophysics (CAASTRO), Sydney Institute of Astronomy, University of Sydney, Australia.

[18] Max Planck Institute for Astrophysics, Karl Schwarzschild Str. 1, 85741 Garching, Germany.

[19] University of Twente, P.O. Box 217, 7500 AE Enschede, The Netherlands

[20] Harvard-Smithsonian Center for Astrophysics, 60 Garden Street, Cambridge, MA 02138, USA.

[21] Institute of Astronomy, University of Edinburgh, Royal Observatory of Edinburgh, Blackford Hill, Edinburgh EH9 3HJ, UK.

[22] Leiden Observatory, Leiden University, PO Box 9513, 2300 RA Leiden, The Netherlands.

[23] Hamburger Sternwarte, University of Hamburg, Gojenbergsweg 112, 21029 Hamburg, Germany.

[24] Leibniz-Institut für Astrophysik Potsdam (AIP), An der Sternwarte 16, 14482 Potsdam, Germany.

[25] Research School of Astronomy and Astrophysics, Australian National University, Mt Stromlo Obs., via Cotter Road, Weston, A.C.T. 2611, Australia.

[26] Thüringer Landessternwarte, Sternwarte 5, D-07778 Tautenburg, Germany

[27] Department of Astrophysics/IMAPP, Radboud University Nijmegen, P.O. Box 9010, 6500 GL Nijmegen, The Netherlands.

[28] Laboratoire Lagrange, UMR7293, Université de Nice Sophia-Antipolis, CNRS, Observatoire de la Côte d'Azur, 06300 Nice, France.

[29] International Centre for Radio Astronomy Research - Curtin University, GPO Box U1987, Perth, WA 6845, Australia.

[30] Center for Information Technology (CIT), University of Groningen, The Netherlands.

[31] Centre de Recherche Astrophysique de Lyon, Observatoire de Lyon, 9 av Charles Andr\'{e}, 69561 Saint Genis Laval Cedex, France.

[32] National Radio Astronomy Observatory, 520 Edgemont Road, Charlottesville, VA 22903-2475, USA.

[33] LESIA-Observatoire de Paris, CNRS, UPMC Univ. Paris-Diderot, 92190   Meudon, France.

[34]Department of Physics and Electronics, Rhodes University, PO Box 94, Grahamstown, 6140, South Africa

[35]SKA South Africa, 3rd Floor, The Park, Park Road, Pinelands, 7405, South Africa

[36]Argelander-Institut für Astronomie, University of Bonn, Auf dem Hügel 71, 53121, Bonn, Germany.

*To whom correspondence should be addressed. E-mail: W.Hermsen@sron.nl

**Abstract**: Pulsars emit low-frequency radio waves through to high-energy gamma-rays that are generated anywhere from the surface out to the edges of the magnetosphere. Detecting correlated mode changes in the multi-wavelength emission is therefore key to understanding the physical relationship between these emission sites. Through simultaneous observations, we have detected synchronous switching in the radio and X-ray emission properties of PSR B0943+10. When the pulsar is in a sustained radio 'bright' mode, the X-rays show only an un-pulsed, non-thermal component. Conversely, when the pulsar is in a radio 'quiet' mode, the X-ray luminosity more than doubles and a 100%-pulsed thermal component is observed along with the non-thermal component. This indicates rapid, global changes to the conditions in the magnetosphere, which challenge all proposed pulsar emission theories.

**Main Text:**
Radio pulsars are powered by the energy released as the highly magnetized neutron star spins down. The radio pulses are generated in the pulsar magnetosphere, most probably close to the neutron star surface (*1,2*). Shortly after the discovery of pulsars, it was observed that the radio pulse behavior can discretely change on timescales as short as a rotation period. These changes in emission mode can manifest as switches between ordered and disordered states or variations in intensity and pulse shape, including the complete cessation of observable radio emission (*3,4*).

Because the emitted radio luminosity is a negligible fraction of the available spin-down energy, usually substantially less than $10^{-5}$, this phenomenology was presumed to be related solely to microphysics of the radio emission mechanism itself. This perception has recently been challenged by the identification of a relationship between the spin properties of neutron stars and their radio emission modes. PSR B1931+24 was observed to cease emitting for tens of days, during which it spins down ~50% less rapidly (*5*). PSR J1841–0500 (*6*) and PSR J1832+0029 (*7*) exhibit similar behaviors. A number of other pulsars display smaller changes in spin-down rate, which correlate with variations in their average radio pulse shapes (*8*). The implication of these results is that mode changing is due to an inherent, perhaps universal pulsar process which causes a sudden change in the rate of angular momentum loss that is communicated along the open field lines of the magnetosphere. Whereas changes in spin-down rate can only be detected on time-scales of a few days or longer, the recently identified link with the rapid switching observed in radio emission modes suggests a transformation of the global magnetospheric state in less than a rotation period. Despite the recent flurry of pulsar detections at high energies (*9*), the only causal relation between the radio pulses and emission at other wavelengths, likely emanating from different locations in the

magnetosphere, has been made for optical emission and giant radio pulses from the Crab pulsar (10)

PSR B0943+10 is a paragon of mode-changing pulsars. Relatively old (characteristic age 5 Myr), with a long spin period (P = 1.1 s), it switches at intervals of several hours between a radio-bright, highly organized mode (B) and a quieter chaotic mode (Q) (*11,12*). At the B- to Q-mode transition a sub-pulse drifting structure dissolves within a few seconds, the emission becomes disorsganized, and an additional highly polarized radio component appears, preceding the main pulse by 52° of pulse longitude (*13*). PSR B0943+10 has also been detected in two short observations with the XMM-Newton observatory as a weak X-ray source (*14*). Assuming the X-rays to be thermal, this was used to support a model in which one system of streaming particles produces both the sub-pulse-modulated radio emission directly, as well as thermal X-ray emission through bombardment of the polar cap surface (*15*)**.** Because the particle streams are thought to be determined by the magnetosphere as a whole, detection of simultaneous X-ray/radio mode switching would uniquely probe the interaction between local and global electromagnetic behavior. Moreover, this would strengthen the earlier conclusion that entire magnetospheres change, settling down within a few seconds.

To test these hypotheses, we carried out a simultaneous X-ray/radio observing campaign on PSR B0943+10, between 4 November and 4 December 2011. These observations were designed to investigate what changes, if any, occurred in the X-rays when the radio emission changed mode. The X-ray observations consisted of six 6-hr observations in the 0.2 - 10 keV energy band with ESA's XMM-Newton space observatory ((*16*), Table S1), accompanied by radio observations with the Indian Giant Metrewave Radio Telescope (GMRT) at 320 MHz and the international Low Frequency Array (LOFAR) at 140 MHz, both simultaneously.

To identify the radio B- and Q-mode time windows, the radio pulse sequences were folded with up-to-date ephemerides from the Jodrell Bank long-term timing program (*17*) (Fig. 1). We could determine the times of mode switches from GMRT and LOFAR data with an accuracy of a few seconds. Table S2 lists the used B- and Q-mode time windows, which completely cover our XMM-Newton observations. In the ~30 hrs of usable X-ray observations, PSR B0943+10 spent roughly equal amounts of time in the B- and Q-modes.

PSR B0943+10 was clearly detected in each of our XMM-Newton observations with the simultaneously used CCD detectors PN (*18*) and MOS-1+2 (*19*) of the European Photon Imaging Camera (EPIC). The derived count rates ranged from that of the previously reported value for the PN detector of $(0.38 \pm 0.07) \times 10^{-2}$ counts/s (0.5-8 keV) (*14*) up to about twice that value, providing evidence for X-ray variability in an old, rotation-powered pulsar. Dividing the 0.2-10 keV X-ray events into the radio-derived B- and Q-mode time windows, we found the X-ray count rate to be more than a factor of two higher in the radio Q-mode than in the B-mode (Fig. S1). In the B-mode, the PN CCDs had a count rate of $(0.44 \pm 0.07) \times 10^{-2}$ counts/s, whereas in the Q-mode this more than doubled to $(1.08 \pm 0.08) \times 10^{-2}$ counts/s. This was independently confirmed with the MOS detectors, providing evidence for simultaneous mode switching in the radio and X-ray properties.

To search for X-ray pulsations, we selected events in the PN/MOS-1+2 CCDs that arrived in the Q-mode time window and within a radius of 15 arcseconds from the source position. From this, we obtained a 6.6-σ detection of a pulsed signal (top panel of Fig. 2B) at a period consistent with the rotational frequency predicted by the Jodrell Bank ephemeris (Table S3). The pulse profile (energies 0.5-2 keV) is broad. Surprisingly, the X-ray events detected during the radio B-mode do not show any evidence for a pulsed signal (top panel Fig. 2A). Figure 2 shows that the broad X-ray pulse in the Q-mode covers the phases of the main radio pulse and precursor; the latter is clearly visible in the Q-mode, 52° (0.14 phase) ahead of the main pulse at 320 MHz (Fig. 2B).

X-ray spectral analysis (*16*) reveals two components in the Q-mode . The best spectral fit to the total (i.e. pulsed and unpulsed) spectrum is the sum of a power-law and thermal black-body component (Fig. 3A with fit parameters in Tables 1 and S4). The spectrum of the pulsed component in the Q-mode is best described by a single thermal black-body model (Fig. 3B, Tables 1 and S4). It appears that the spectral fits to the thermal component in the total Q-mode spectrum and the thermal pulsed spectrum in the Q-mode are statistically consistent ($\Delta$ flux is 0.1 $\sigma$ and $\Delta kT=2.5$ $\sigma$). This means that the Q-mode total X-ray emission consists of an un-pulsed component with a steep, non-thermal power-law spectrum, and a ~100% pulsed component with a thermal black-body spectrum. This is also reflected in the variation of the pulsed fraction with energy (Table S5). In the B-mode, the spectrum can be satisfactorily described with a single power-law as well as a single black-body shape (Table S4). However, the most likely shape is a non-thermal spectrum (Fig. 3C), indistinguishable from the non-thermal component in the total Q-mode spectrum (supplementary online text).

PSR B0943+10 is one of just 10 old (characteristic age > 1Myr), non-recycled radio pulsars where X-ray emission has also been detected (*20-22*). Although the surfaces of such pulsars have cooled substantially since birth, the observed X-ray emission is argued to be thermal in some cases. This has led to the conclusion that such pulsars may have 'hotspots' on their magnetic polar caps, generated by the bombardment of particles accelerated in the radio emission process. In all models, the bombarding particles result from pair-creation in the pulsar magnetosphere. In polar-vacuum-gap models (*1,23*) this occurs directly above the surface. Recent adaptations of the model (*15*) predict the thermal X-ray brightness of pulsars that exhibit regular modulation of their radio sub-pulse drift. Such modulation is indeed observed in the B-mode of B0943+10 (*24*), and the original X-ray detection of this pulsar (*14*) appeared to support this prediction, if one assumes the X-rays have a thermal origin. However, while the reported count rate suggests that the pulsar was in the B-mode during those observations, our spectral analysis shows that the X-ray emission in the B-mode is actually non-thermal. Surprisingly, strong thermal X-rays are only detected in the Q-mode, where the observed radio emission is weak and chaotic.

Space-charge-limited-flow models (*2*) also feature pair-created particles that heat the polar cap via backflow. These differ from polar-gap models, however, in that charged primary particles are freely drawn from the neutron star surface. The thermal luminosities predicted for older pulsars (*25*) fall below what is observed in PSR B0943+10's Q-mode and below the model's expected non-thermal cascade emission, consisting of contributions from curvature radiation, inverse Compton scattering and synchrotron radiation (*26*). Such non-thermal emission, expected on open field-lines, would almost certainly be mode-dependent, in contrast to what we observe.

Modeling of PSR B0943+10's geometry (*27*) strongly argues that the magnetic and spin axes are nearly aligned, with our line-of-sight passing near the pole (*24*) (Fig. 4). This implies that the isotropically emitted X-rays of a polar hotspot should appear un-modulated throughout the rotation period, contrary to the observed Q-mode X-ray pulsations. One possible interpretatrion is then that the observed X-ray pulsations result from time-dependent scattering of the emission within the closed magnetosphere ( Fig. 4), a scenario similar to that proposed to explain magnetospheric eclipses of pulsed radio emission in the double pulsar system PSR J0737–3039 (*28,29*). If so, then the observed thermal emission in the Q-mode is only about half the actual hotspot emission, doubling the inferred area of the hotspot.

If scattering plays an important role in the Q-mode, the absence of X-ray pulsations and thermal X-rays in the B-mode may be attributed to increased scattering. As suggested to explain nulling and mode switching in radio pulsars (*30*), an expansion in the volume of the closed magnetosphere might accompany the mode-change, though it is unlikely to achieve the required degree of screening. It thus appears that at B-mode onset the surface hotspot emission is reduced to undetectable levels within a few seconds, induced by a drastic reduction in the downward flow of charged particles.

In the context of the polar-gap model, it has been suggested (*31*) that mode changes occur when the local surface temperature crosses a critical value, $\sim 10^6$ K, that separates dominant emission mechanisms. This would require hotspots to be detectable in both modes, in contradiction with our results. B-mode emission may represent a cooler mode where curvature radiation dominates. However, the temperature transition remains unexplained and our results strongly suggest that a global, rather than local, mechanism is required. Indeed, for a near-aligned pulsar such as PSR B0943+10, a range of quasi-stable magnetospheric configurations is expected (*32,33*), and the non-linear system is proposed to suddenly switch between specific states, each having a specific emission beam and spin-down rate (*30*).

Whatever the true nature of the mode switch, the contrast between the Q-mode's *enhanced* X-ray emission and *reduced* radio emission may well be illusory. Radio emission is only sampled on field-lines instantaneously directed towards us, and our sightline (*34*) makes only a grazing traverse of the polar cap. The core region of this polar cap, the probable site of X-ray hot spots, remains invisible to us at radio frequencies and may well change differently.

The totality of the mode transition in PSR B0943+10 – changes in radio sub-pulse behavior and profile, the appearance of a precursor and, now, the switching on and off in X-rays of a likely hotspot – implies that we are dealing with a rapid and global magnetospheric state change. Through radio and X-ray `before' and `after' snapshots, we have shown that a magnetosphere ten times the size of Earth completely changes personality within a few seconds, a near-instantaneous transformation that challenges our current understanding of pulsars and magnetospheres in general.

**Supplementary Materials**
www.sciencemag.org
Materials and Methods
Figs. S1, S2
Tables S1, S2, S3, S4, S5
References (35–40) [Note: The numbers refer to any additional references cited only within the Supplementary Materials]

**Acknowledgments:**

We thank the staff of XMM-Newton, the GMRT and LOFAR for making these observations possible. XMM-Newton is an ESA science mission with instruments and contributions directly funded by ESA Member States and the USA (NASA). GMRT is run by the National Centre for Radio Astrophysics of the Tata Institute of Fundamental Research. LOFAR, the Low Frequency Array designed and constructed by ASTRON, has facilities in several countries, that are owned by various parties (each with their own funding sources), and that are collectively operated by the International LOFAR Telescope (ILT) foundation under a joint scientific policy. ASTRON and SRON are supported financially by NWO, the Netherlands Organization for Scientific Research. Two of us (JMR, GAEW) thank the NWO and ASTRON for their Visitor Grants. The used X-ray data can be retrieved from the XMM-Newton Science Archive at xmm.esac.esa.int/xsa. The applied B- and Q-mode window selections, derived from the GMRT and LOFAR observations, are listed in Table S2 of the SOM.

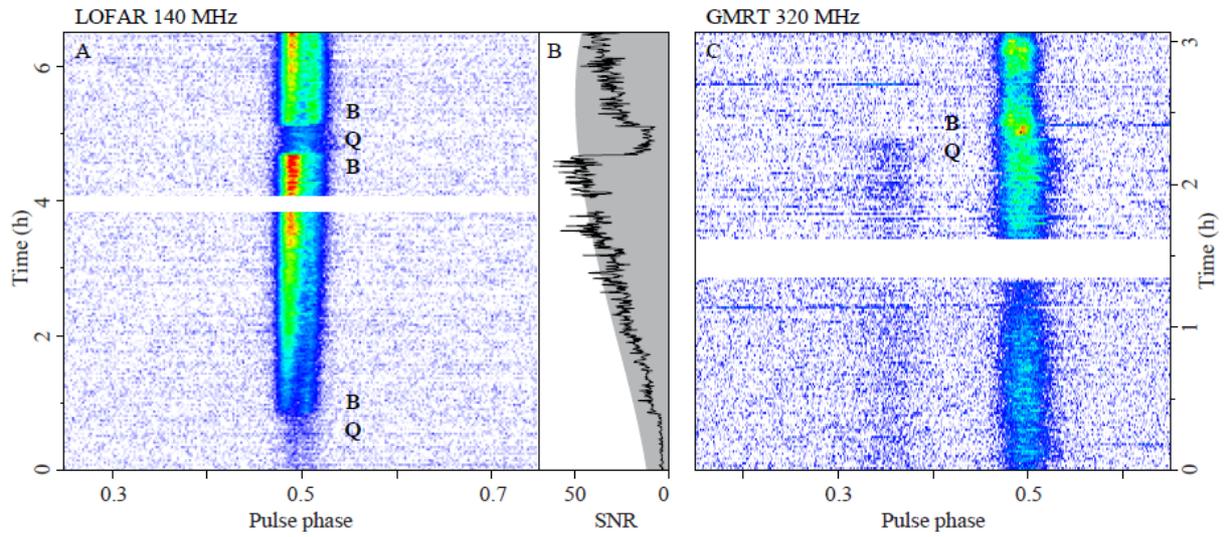

**Fig. 1**. Panel A: identification of the B- and Q-modes with LOFAR at 140 MHz, during XMM-Newton observation #1, showing the pulse intensity versus rotational phase and time. A 10-minute section (at the 4h mark) contaminated by interference is blanked out. Panel B: the measured signal-to-noise ratio compared with the nominal relative LOFAR sensitivity, changing with elevation throughout the observation (gray scale), normalized over the 4.0-4.5h range. Panel C: GMRT detection at 320 MHz of a Q- to B-mode transition in XMM-Newton observation #5. Color scale optimized to show the simultaneous disappearance of the precursor pulse at phase ~0.35. A 15-minute section (at the 1.5h mark) used for rephasing on a continuum source is blanked out.

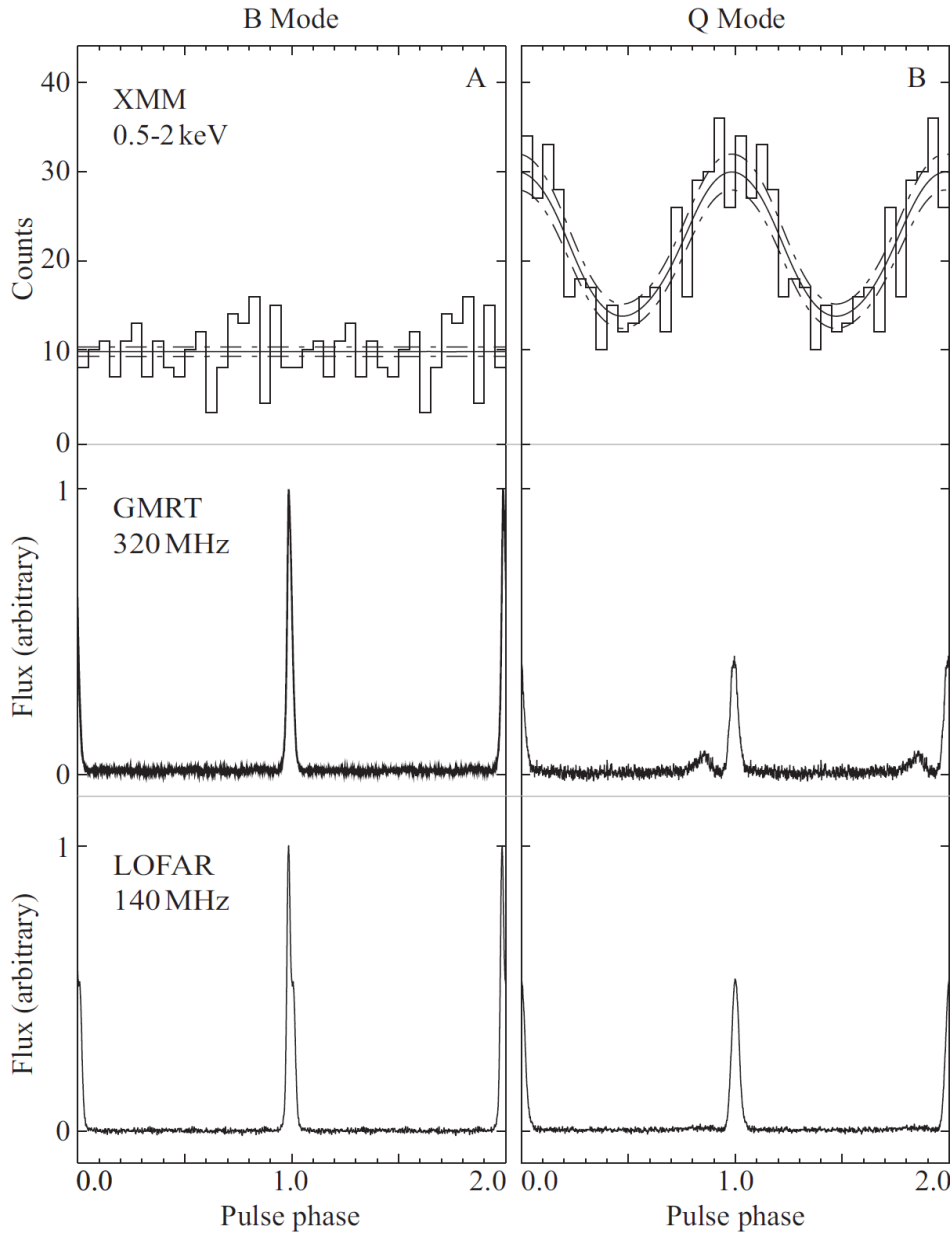

**Fig. 2.** Aligned X-ray and radio pulse profiles of PSR B0943+10 in its B- and Q-modes. Panel A: there is no evidence for a pulsed signal in the B-mode X-ray data, the flat distribution showing constant emission from the pulsar. Panel B: the X-ray profile in the Q-mode represents a 6.6-σ detection on top of a flat constant level. The solid and dotted lines in the X-ray profiles are the Kernel-density estimator and ± 1-σ levels. The weak precursor, present only in the Q-mode, is clearly visible in the GMRT radio profile at 320 MHz at 52° (0.14 phase) prior to the main pulse, and verified to be also weakly present in the LOFAR Q-mode profile.

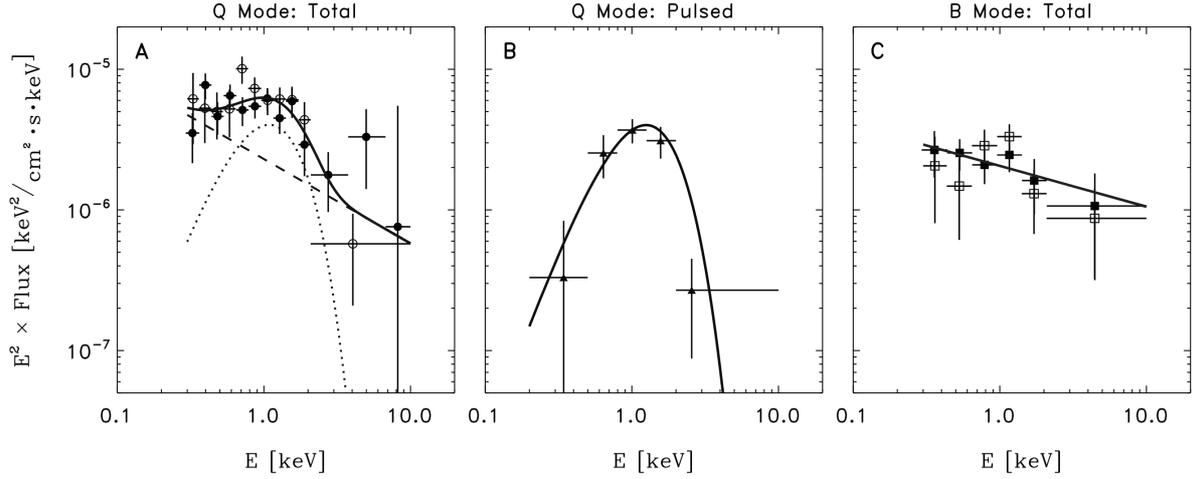

**Fig. 3.** Unabsorbed, i.e. corrected for absorption by interstellar gas along the line-of-sight, X-ray photon spectra of PSR B0943+10. Panel A: Total (i.e. pulsed and unpulsed) spectrum from spatial analyses of skymaps of Q-mode events from the XMM EPIC PN CCDs (filled symbols) and MOS1+2 CCDs (open symbols). The broken line shows the power-law component, and the dotted line the black-body component of the best fit. Panel B: The spectrum of the pulsed emission detected only in the Q-mode, analyzing the pulse profiles measured with the PN and MOS1+2 CCDs. The solid curve shows the best black-body fit. Panel C: The total spectrum, as in panel A, but here for the B-mode time windows. The solid line shows the best-fit power-law spectrum. The thermal black-body components in panels A and B are statistically the same; the non-thermal power-law components in panels A and C are also fully consistent with being identical. All error bars are 1 $\sigma$. For fit parameters see Table 1.

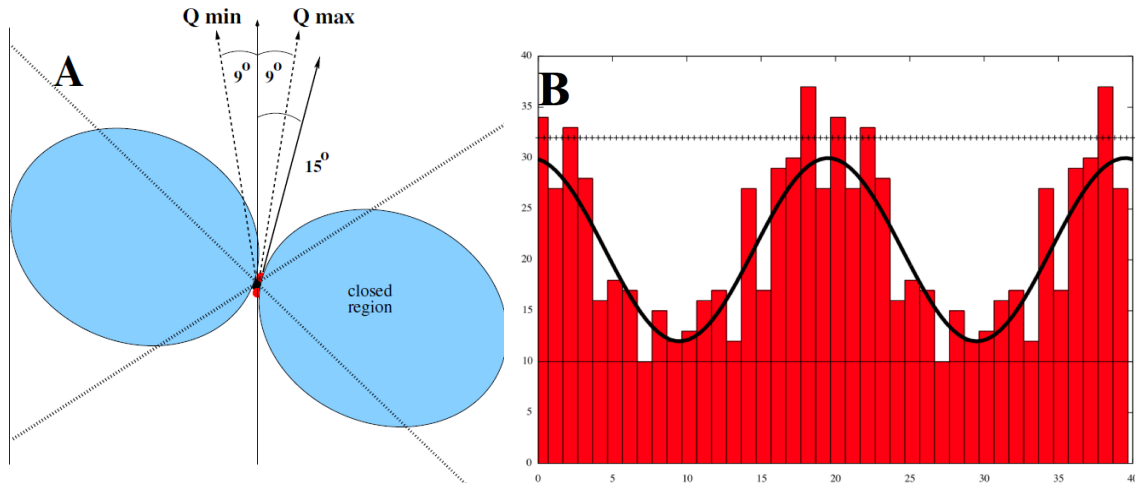

**Fig. 4.** Panel A. The geometry of PSR B0943+10. The dipole magnetic axis is inclined at 15° to the (vertical) rotation axis and the diagonal lines indicate the location of the null-charge surface, separating regions of positive and negative charge. As the pulsar rotates the observer's line of sight maintains an angle of 9° to the rotation axis, with Q*max* indicating the alignment when the radio pulse is seen (in both modes). The blue region indicates the supposed closed region, bounded by the light cylinder (at which the co-rotating magnetosphere reaches a rotation velocity equal to the speed of light) at 52,000 km. If, in a toy model, extinction of thermal X-rays from the polar cap is proportional to the extent of the traversed closed region (≈ 10,000 km at Q*min* and ≈ 500 km at Q*max*) then we obtain a sinusoidal fit to the observed Q-mode X-ray pulse (Panel B) so that maximum recorded emission corresponds to minimum extinction and vice versa. The lower horizontal line corresponds to the level of steady non-thermal emission and demonstrates that extinction of the thermal emission is near total at Q*min*. The upper horizontal line then represents the level of actual unscattered X-ray emission from the polar cap.

**Table 1.** Spectral parameters for the best model fits to the X-ray spectra shown in Fig. 3. Fits are made with a black-body (BB) shape and/or a power-law (PL) shape for the total and the pulsed emissions in the Q-mode window and the total emission in the B-mode. The column density $N_H$ has been fixed at $4.3 \times 10^{20}$ cm$^{-2}$.

| Mode total / pulsed | Model | BB (kT) keV | PL index $\Gamma$ ($\alpha E^{-\Gamma}$) | BB flux, unabs $10^{-15}$ erg cm$^{-2}$ s$^{-1}$ | PL flux, unabs $10^{-15}$ erg cm$^{-2}$ s$^{-1}$ | $\chi^2_{red}$ / dof |
|---|---|---|---|---|---|---|
| **Q total** | BB+PL | 0.277±0.012 | 2.60±0.34 | 7.52±2.20 | 7.55±1.81 | 0.81 / 20 |
| **Q pulsed** | BB | 0.319±0.012 | | 7.81±1.64 | | 0.38 / 3 |
| **B total** | PL | | 2.29±0.16 | | 7.69±1.00 | 0.74 / 10 |

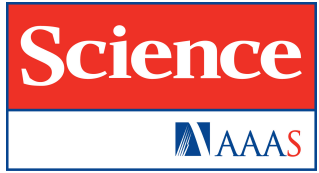

# Supplementary Materials for

## Synchronous X-ray and Radio Mode Switches: a Rapid Global Transformation of the Pulsar Magnetosphere

W. Hermsen[*], J.W.T. Hessels, L. Kuiper, J. van Leeuwen, D. Mitra, J. de Plaa, J. M. Rankin, B. W. Stappers, G.A.E. Wright, R. Basu, A. Alexov, T. Coenen, J.-M. Grießmeier, T.E. Hassall, A. Karastergiou, E. Keane, V.I. Kondratiev, M. Kramer, M. Kuniyoshi, A. Noutsos, M. Serylak, M. Pilia, C. Sobey, P. Weltevrede, K. Zagkouris, A. Asgekar, I.M. Avruch, F. Batejat, M.E. Bell, M.R. Bell, M.J. Bentum, G. Bernardi, P. Best, L. Bîrzan, A. Bonafede, F. Breitling, J. Broderick, M. Brüggen, H.R. Butcher, B. Ciardi, S. Duscha, J. Eislöffel, H. Falcke, R. Fender, C. Ferrari, W. Frieswijk, M.A. Garrett, F. de Gasperin, E. de Geus, A.W. Gunst, G. Heald, M. Hoeft, A. Horneffer, M. Iacobelli, G. Kuper, P. Maat, G. Macario, S. Markoff, J.P. McKean, M. Mevius, J.C.A. Miller-Jones, R. Morganti, H. Munk, E. Orrú, H. Paas, M. Pandey-Pommier, V.N. Pandey, R. Pizzo, A.G. Polatidis, S. Rawlings, W. Reich, H. Röttgering, A.M.M. Scaife, A. Schoenmakers, A. Shulevski, J. Sluman, M. Steinmetz, M. Tagger, Y. Tang, C. Tasse, S. ter Veen, R. Vermeulen, R.H. van de Brink, R.J. van Weeren, R.A.M.J. Wijers, M.W. Wise, O. Wucknitz, S. Yatawatta, P. Zarka

*To whom correspondence should be addressed. E-mail: W.Hermsen@sron.nl

**This PDF file includes:**

Materials and Methods
References
Figs. S1 to S2
Tables S1 to S5

**Materials and Methods**

**XMM-Newton observations**

Between 4 November and 4 December 2011 we were awarded six XMM-Newton observations (identifiers #1-#6) of about 6 hours duration each in the energy band 0.2 - 10 keV. Table S1 lists the effective exposure times and instrumental setup of the simultaneously used CCD detectors of the European Photon Imaging Camera (EPIC): the PN (*18*) and the MOS-1+2 detectors (*19*). These observations were performed simultaneously with the GMRT at 320 MHz and LOFAR at 140 MHz, firstly to guarantee that we would recognize all abrupt (on time scales of seconds) radio mode changes, but also to be able to study the timing, spectral and polarization characteristics of PSR B0943+10 during the X-ray observations. We did not use the last XMM-Newton observation, which suffered from many solar soft proton flares, resulting in a high background.

**GMRT setup and mode determinations**

The observations with the GMRT, located near Pune in India, were conducted in the same way as an earlier 8-hour session (*35*) using 20 of the 30 antennas, 14 in the central square and 2 each in the three arms, for ease and stability of the phasing. Short interruptions of the PSR B0943+10 observations were needed every 2-3 hours for re-phasing on a continuum radio source. The array was first phased on the flux calibrator 3C147 and then slewed to a phase calibrator 0837–198 and 1021+216 close to PSR B0943+10 to check for phase stability and re-phasing as required. A bandwidth of 33.33 MHz was used (limits 306 - 339.33 MHz) and the data were recorded using the pulsar mode of the GMRT software backend (*36*) at a time resolution of 0.98304 ms split into 256 channels. After de-dispersion, sections of the data with baseline variations due to broad band interference were mitigated using running mean subtraction as well as flattening of the baseline on a pulse-to-pulse basis. Displays of the processed observations in short averages then provided the needed modal transition times. Figure 1, panel C (in main text) shows an example of a Q- to B-mode transition measured with the GMRT at 320 MHz. In this figure, the disappearance of the weak precursor in the B-mode after the transition is clearly visible.

**LOFAR setup and mode determinations**

The LOFAR data were acquired by combining the 12 most centrally located high-band antenna sub-stations (the 'Superterp' HBAs, Fig. 1 in (*37*)) located in Exloo, The Netherlands. These 12 sub-stations receive the same clock signal, which greatly facilitates the coherent, 'in phase', addition of their signals. A coherent beam was formed on the position of PSR B0943+10 and tracked throughout the observation. A second beam was formed on nearby pulsar PSR B0950+08 to serve as a reference beam for system sensitivity and radio frequency interference checks. For each beam, the dual-polarization array signal was transformed on-line to the four Stokes parameters. These were sampled every 0.32 ms, for a total bandwidth of 46.875 MHz split into 3840 channels (*38*).

For each of the 6 epochs, the data were folded with an ephemeris from the Jodrell Bank long-term timing program (*17*). Per 1-minute sub-integration, the profile signal-to-noise ratio was calculated versus time, and compared to the expected LOFAR sensitivity curve. Simultaneously, the pulse profile was plotted versus time and phase (Fig. 1, Panel A). In the intensity plots, mode changes were initially identified by eye. We could see the times of mode switches with an accuracy of a few seconds. However, we determined in 10-second resolution plots the UT times for all mode changes generally to 1-minute accuracy, sufficiently accurate for the selection of the X-ray data. The visual inspection of the intensity plots, and the coincidence testing between the PSR B0943+10 and PSR B0950+08 data, allowed for the flagging of two ~5-min sections of broadband radio frequency interference in 42 hours of data.

Table S2 lists time windows used in this work. In the useful XMM-Newton observations (#1 - #5) the B- and Q-mode each appeared for roughly equal amounts of time.

**X-ray spatial analysis**

Even though PSR B0943+10 is a weak X-ray source (*14*), our long observations detected it easily. In a detailed spatial analysis, we produced maximum-likelihood-ratio maps, which give in each sky bin the ratio of the likelihood that a source is located in that bin in addition to background (in this case assumed to be a flat background level) over the likelihood that just the background (no source) is present, a test statistic for the presence of a source (3*9*). With this method we clearly detected PSR B0943+10 in each of the five XMM-Newton observations using events from the PN detector, and consistently when selecting events from the MOS-1+2 detectors. Using the sensitive area arfgen-1802 and redistribution matrix rmfgen-1561 (implemented in XMMSAS-20110223-1801-1100 (*40*)) for the XMM-response, the derived flux values ranged from the previously published value (*14*) up to more than twice that value, providing significant evidence for time variability. In the next step we separately analyzed events that arrived in the radio B- and Q-mode time windows (Table S2). The detection significance for a similar exposure is twice as high in the radio Q-mode as in the B-mode, and the count rate is more than a factor two higher in the Q-mode than in the B-mode, shown in Fig. S1 for the PN detector. This provides us with evidence for simultaneous mode switching in the radio and X-ray bands.

**X-ray timing analysis.**

We searched for pulsations in the X-ray data for events detected during the radio Q-mode, in which we had more than twice the number of source counts detected in the B-mode. In order to maximize our statistics, we selected all events within a radius of 15" from the source position (see Fig. S2) for the PN, MOS-1+2 detectors. We verified that this radius gives the optimum signal-to-noise ratio for the XMM point-spread functions. The event arrival times at the satellite have been converted to arrival times at the Solar System Barycenter adopting our new accurate X-ray source position which has a negligible statistical error and a systematic uncertainty of at most 1". Folding these event arrival times using an updated Jodrell Bank ephemeris (fixed X-ray source location, see Table S3) yielded a 6.6-σ signal adopting the $Z_1^2$ test (Rayleigh test) and revealed a broad pulse profile (see top panel Fig. 2B). The profile can be fitted equally well with a sinusoid or a flat background with a Gaussian shaped pulse. The latter Gaussian pulse template reaches its maximum at phase 0.98 ± 0.02 with σ = 0.20 ± 0.05 and is an excellent fit to the 0.5 - 2 keV pulse shown in Fig. 2. We found that X-ray pulsations were detected significantly only in the 0.5-2 keV energy band: 1.2 σ for 0.2-0.5 keV and 1.9 σ for 2-10 keV.

The X-ray events detected during the radio B-mode do not show any evidence for a pulsed signal (1.2 σ for the integral 0.2-10 keV), as can also be seen in the top panel of Fig. 3A. This is an additional manifestation of X-ray mode switching in a radio pulsar, this time revealed in its X-ray timing properties.

**X-ray spectral analyses**

We first present the results for the total (i.e. pulsed and un-pulsed) X-ray emission in the X-ray-bright, radio-quiet Q-mode. We again use the maximum-likelihood analysis, applying this to the two-dimensional counts sky maps in different energy bins measured with the EPIC PN detector and separately to the summed counts maps for EPIC MOS-1+2. The latter maps have only slightly lower statistics than the PN counts maps and give independent results. The maximum likelihood analysis delivers a count(-rate) spectrum, which can be de-convolved assuming different source input spectra in model fitting. An additional important parameter is the hydrogen column density in the source direction ($N_H$), causing significant absorption below 2 keV. We fixed $N_H$ at $4.3 \times 10^{20}$ cm$^{-2}$ for all model fits, the value used in the first study of PSR B0943+10 (*14*). We verified that $N_H$ values reported in the literature, differing by up to ~35 %, influenced the model-fit parameters only within their 1σ uncertainties.

Figure 3A shows the unabsorbed (i.e. corrected for absorption by the hydrogen column density) total X-ray photon spectrum in the radio Q-mode. The filled symbols are the flux values for the PN data and the open symbols for the MOS-1+2 data. The best model fitted to both sets of flux values simultaneously is the sum of a component with a power-law shape (broken line) and one with a BB-shape (dotted line). This fit was an improvement at the 4.1 σ level with respect to a fit with a single PL, and 3.5 σ relative to a single BB-model fit. All spectral parameters for the three model fits (BB+PL, BB, PL) are given in Table S4. The contributions of the two components in the X-ray band 0.5 - 8 keV are equal, each $\sim 7.5 \times 10^{15}$ erg cm$^{-2}$ s$^{-1}$, to be compared with the value of $4.4^{+1.8}_{-1.5} \times 10^{-15}$ erg cm$^{-2}$ s$^{-1}$ reported in the first detection (*14*) for a PL fit to the total spectrum with the same index $\Gamma = 2.6^{+0.7}_{-0.5}$. This suggests that in this early observation of 20-ks duration PSR B0943+10 was predominantly in the B-mode, when the X-ray flux is low.

The spectrum of the pulsed component in the Q-mode can be determined by first deriving the pulse excess counts above a flat background level in the pulse profiles for differential energy bins. These excess counts were derived by fitting with the template Gaussian profile derived in the 0.5 - 2 keV energy band (parameters given above). We verified that there is no significant variation in profile shape over the total X-ray band. A single BB model gave an excellent fit, a PL model was acceptable only at the 3% level (main discrepancy at 0.2 - 0.5 keV energies; see also in Table S5 the lack of evidence for pulsed emission below 0.5 keV). Figure 3B shows the reconstructed unabsorbed photon spectrum of the pulsed component with the best fit model superposed. Full details of the two model (BB and PL) fits are given in Table S4.

The spectral parameters of the thermal component in the best fit (PL + BB) to the total Q-mode spectrum and the thermal pulsed spectrum in the Q-mode appear to be statistically consistent (Δ flux is 0.1 σ and ΔkT=2.5 σ). In other words, this means that the Q-mode total X-ray emission consists of an un-pulsed component with a non-thermal PL spectrum, and a pulsed component with a thermal BB spectrum. This is also reflected in the variation of the pulsed fraction with energy (see Table S5). It is important to note that XMM-Newton is very sensitive in the low-energy interval 0.2-0.5 keV, and PSR B0943+10 is detected at a high significance in the sky maps in this energy band, e.g. in just the PN sky map at a level of 12 σ. Therefore, the non-detection of a pulsed signal below 0.5 keV is not due to a lack of sensitivity of XMM-Newton.

Finally, the total X-ray spectrum in the radio B-mode can be derived in analyses of counts sky maps as explained above for the total spectrum in the Q-mode. In this case the counting statistics are low, and it is not possible to discriminate between a single PL and a BB model, both give good fits (Table S4 and Fig. S2). Figure 3C shows the un-pulsed total B-mode spectrum assuming a PL spectrum. We selected this fit for the composite Fig.3, because we find it indicative that the total un-pulsed X-ray emission in the radio B-mode has a spectrum statistically identical to the un-pulsed PL component in the Q-mode (Δ flux is 0.05 σ and ΔΓ is 0.8 σ). This would lead to the conclusion that the mode switching from B- to Q-mode means the addition of a new thermal component consistent with 100% pulsation in the X-ray domain, and no change in the un-pulsed non-thermal component. If one would assume on the other hand, the B-mode un-pulsed X-ray spectrum to be thermal (BB with kT = 0.250±0.006 keV, and flux = 5.36±0.78) $10^{-15}$ erg cm$^{-2}$s$^{-1}$, see Table S4), then in the mode switching the un-pulsed thermal component in the B-mode has to be replaced in the Q-mode by an un-pulsed power-law component and, in addition, a new pulsed thermal component has to be added. We consider the first interpretation of the B-mode spectrum most likely.

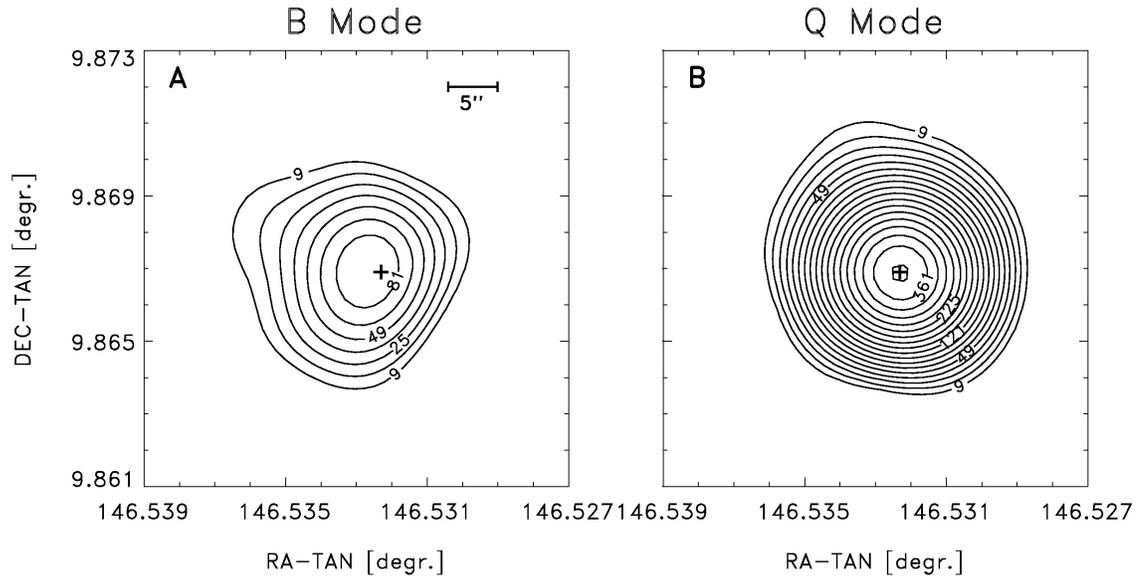

**Fig. S1.** Maximum-likelihood-ratio sky maps for X-ray energies 0.2-10 keV showing evidence for the detection of PSR B0943+10 with the PN CCDs aboard XMM-Newton in the two radio-defined modes. Panel A: for events selected in the radio B-mode time windows a 9.9 σ detection significance and count rate $(0.44 \pm 0.07) \times 10^{-2}$ counts/s. Panel B: for events in the radio Q-mode a 20 σ detection and more than twice the count rate $(1.08 \pm 0.08)\ 10^{-2}$ counts/s. The contour values give the variance of the detection significances in number of σ's, assuming one degree of freedom. The + symbol gives the best location for PSR B0943+10 derived from these XMM-Newton observations (see Table S3).

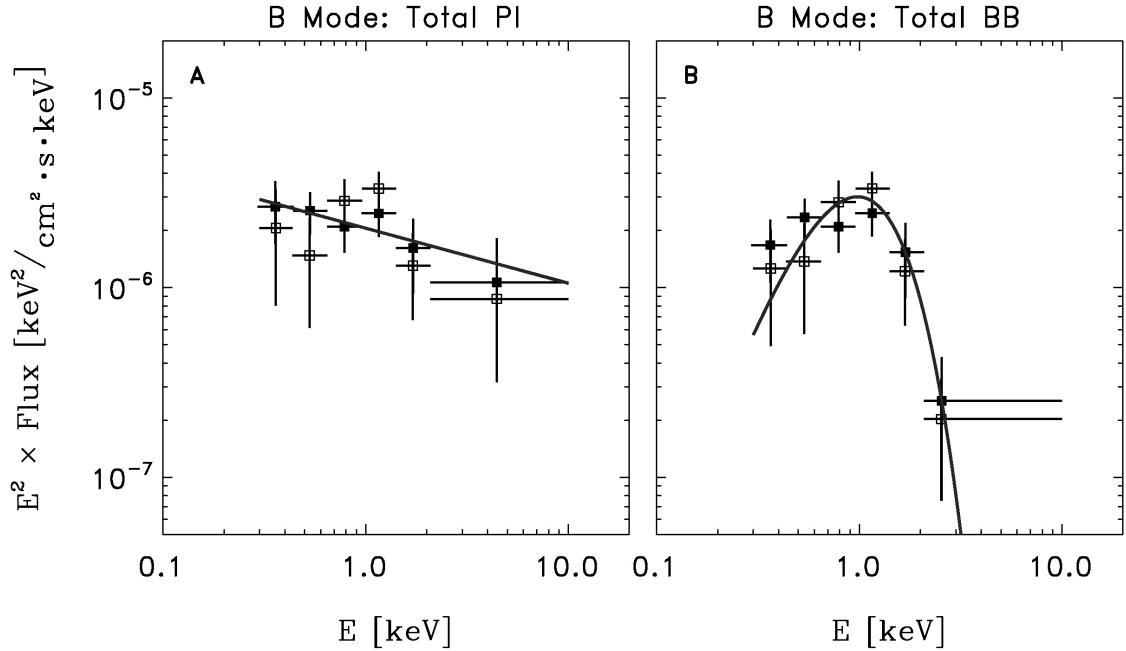

**Fig. S2**. Unabsorbed total (i.e. pulsed and un-pulsed) X-ray photon spectra of PSR B0943+10 from spatial analyses of events selected in the radio B-mode time windows for the XMM EPIC PN CCDs (filled symbols) and MOS1+2 CCDs (open symbols). Panel A: flux values assuming an underlying PL source spectrum, the solid line show the best fit. Panel B: flux values for a BB spectrum, the solid line shows the best thermal fit. Note: the measured count rates are the same for both figures, but the de-convolved flux values depend on the assumed source spectrum. Both spectral shapes give equally good fits to the measured count-rate spectra (see Table S4).

**Table S1.** XMM-Newton effective (corrected for dead time) observation times (kilo seconds) in November and December 2011 for the PN, MOS-1 and MOS-2 CCDs of EPIC. The first row gives the observation identifier and day. The selected detector modes, indicated in the last column, allow spatial and pulsar-timing studies. Observation #6 suffered from too many solar proton flares and has not been used in this work.

| OBSID, date (day/month)/ CCDs | #1 4/11 | #2 6/11 | #3 21-22/11 | #4 27-28/11 | #5 1-2/12 | #6 4/12 | Detector mode |
|---|---|---|---|---|---|---|---|
| EPIC PN | 15.3 | 24.2 | 23.0 | 20.7 | 20.8 | 19.0 | Full Frame |
| EPIC MOS-1 | 21.7 | 25.9 | 24.7 | 22.3 | 22.5 | 20.7 | Small Window |
| EPIC MOS-2 | 21.7 | 25.9 | 24.7 | 22.4 | 22.5 | 20.7 | Small Window |

**Table S2.** B- and Q-mode time windows identified in both our GMRT and LOFAR observations, covering completely our XMM-Newton observations with observation identifiers (OBSID) #1 to #5. The last column marks the UTC start and end times of individual B- and Q- modes with an accuracy of a minute, sufficiently accurate for the X-ray selections.

| XMM OBSID | Radio mode | Day | Start - end UTC |
|---|---|---|---|
| #1 | Q | 2011-11-04 | 00:55 - 01:45 |
|  | B | 2011-11-04 | 01:45 - 05:35 |
|  | Q | 2011-11-04 | 05:35 - 06:03 |
|  | B | 2011-11-04 | 06:03 - 07:25 |
| #2 | Q | 2011-11-06 | 00:55 - 07:10 |
|  | B | 2011-11-06 | 07:10 - 08:35 |
| #3 | Q | 2011-11-21 | 23:30 - 06:10 |
|  | B | 2011-11-22 | 06:10 - 07:00 |
| #4 | Q | 2011-11-27 | 23:30 - 02:40 |
|  | B | 2011-11-28 | 02:40 - 06:15 |
| #5 | Q | 2011-12-01 | 23:00 - 01:53 |
|  | B | 2011-12-02 | 01:53 - 06:00 |

**Table S3.** Jodrell Bank ephemeris of PSR B0943+10 valid during our XMM-Newton observations with the improved source position from our X-ray observations.

| | |
|---|---|
| $\alpha_{2000}$ | $09^h\ 46^m\ 7^s.787$ |
| $\delta_{2000}$ | $09°\ 52'\ 0''.76$ |
| Epoch (TDB) | 54226 |
| $\nu$ | 0.910989538329 Hz |
| $d\nu/dt$ | $-2.94219 \times 10^{-15}$ Hz s$^{-1}$ |
| $d^2\nu/dt^2$ | $-1.39 \times 10^{-25}$ Hz s$^{-2}$ |
| Start (MJD) | 52573 |
| End (MJD) | 55880 |
| Solar System Ephem. | DE200 |

**Table S4.** Spectral parameters with 1σ errors for the best model fits to the X-ray spectra for: i) the radio Q-mode windows of the total emission measured in the sky map, ii) the radio Q-mode of the pulsed emission measured in the pulse profiles, and iii) the radio B-mode windows of the total emission. The model spectra are black body (BB), power law (PL); the normalization of the PL is at 1 keV. The column density $N_H$ has been fixed at $4.3 \times 10^{20}$ cm$^{-2}$. The unabsorbed fluxes (F) and isotropic luminosities (L) for the BB and PL components are calculated for the energy interval 0.5 - 8 keV and a source distance of 630 pc.

| Mode | Q | Q | Q | Q | Q | B | B |
|---|---|---|---|---|---|---|---|
| total / pulsed | total | total | total | pulsed | pulsed | total | total |
| Model | BB+PL | BB | PL | BB | PL | BB | PL |
| $BB_{norm}$ x $10^4$ | 1.44±0.27 | 3.94 ± 0.23 | | 0.81±0.11 | | 1.61±0.16 | |
| BB (kT) keV | 0.277±0.012 | 0.249±0.003 | | 0.319±0.012 | | 0.250±0.006 | |
| $PL_{norm}$ x $10^6$ | 2.30±0.49 | | 4.74±0.27 | | 2.47±0.37 | | 2.06±0.21 |
| PL, Γ ($\alpha E^{-\Gamma}$) | 2.60±0.34 | | 2.44±0.08 | | 1.97±0.18 | | 2.29±0.16 |
| $F_{BB}$ x $10^{15}$ erg cm$^{-2}$ s$^{-1}$ unabsorbed | 7.52±2.20 | 12.88±1.01 | | 7.81±1.64 | | 5.36±0.78 | |
| $L_{BB}$ $10^{29}$ erg s$^{-1}$ | 3.6±1.0 | 6.1±0.5 | | 3.7±0.8 | | 2.5±0.4 | |
| $F_{PL}$ x $10^{15}$ erg cm$^{-2}$ s$^{-1}$ unabsorbed | 7.55±1.81 | | 16.49±1.10 | | 11.21±2.21 | | 7.69±1.00 |
| $L_{PL}$ $10^{29}$ erg s$^{-1}$ | 3.6±0.9 | | 7.9±0.5 | | 5.3±1.1 | | 3.7±0.5 |
| $\chi^2_{red}$ / dof | 0.81 / 20 | 1.44 / 22 | 1.67 / 22 | 0.38 / 3 | 3.17 / 3 | 0.88 / 10 | 0.74 / 10 |

**Table S5.** Pulsed fractions of PSR B0943+10, when the pulsar is in the radio Q-mode, as a function of energy, defined as the ratio of the integrated flux in the pulse profile over the total (i.e. pulsed and un-pulsed) flux of the point source measured in the sky maps. Errors are 1 σ.

| Energy interval keV | Pulsed fraction |
|---|---|
| 0.2 - 0.5 | 0.10 ± 0.15 |
| 0.5 - 0.8 | 0.44 ± 0.16 |
| 0.8 - 1.3 | 0.62 ± 0.14 |
| 1.3 - 2.0 | 0.60 ± 0.17 |
| 2.0 - 10 | 0.72 ± 0.53 |